\documentclass[noshowkeys,prstab,groupedaddress,amsmath,showpacs,twocolumn,footinbib]{revtex4}

\usepackage{color}
\usepackage{supertabular}
\usepackage{graphicx}
\usepackage{dcolumn}
\begin{document}

\title{Using conceptual metaphor and functional grammar to explore how language used in physics affects student learning}

\author{David T.  Brookes}

\affiliation{Department of Physics; Loomis Laboratory of Physics; 1110 West Green St.; Urbana, IL 61801-3080}
\author{Eugenia Etkina}
\affiliation{The Graduate School of Education;  10 Seminary Place; New Brunswick, NJ 08901}

\keywords{physics education; language; quantum mehcanics}
\pacs{01.40.Fk;01.40.Ha;03.65.-w}

\begin{abstract}

This paper introduces a theory about the role of language in learning physics.  The theory is developed in the context of physics students' and physicists' talking and writing about the subject of quantum mechanics.  We found that physicists' language encodes different varieties of analogical models through the use of grammar and conceptual metaphor.  We hypothesize that students categorize concepts into ontological categories based on the grammatical structure of physicists' language.  We also hypothesize that students over-extend and misapply conceptual metaphors in physicists' speech and writing.   Using our theory, we will show how, in some cases, we can explain student difficulties in quantum mechanics as difficulties with language. 

%\newline\newline DOI: 10.1103/ PhysRevSTPER.1.000001

\end{abstract}

%\volumeyear{}
%\volumenumber{}
%\issuenumber{}
%\eid{identifier}
%\received[Received: ]{July 17, 2004}

%\revised[Revised: ]{July 20, 2004}

%\accepted[Accepted: ]{July 21, 2004}

%\published[Published: ]{July 22, 2004}

\startpage{1}

\maketitle

\section{Introduction}

\subsection{Our Starting Point}

The goal of this paper is to present a theoretical framework explaining the role of spoken and written language in physics.  This framework can be used to probe how physicists represent their ideas in language and more importantly, to understand how physics students interpret language they read and hear.  We will use the framework to understand the types of meaning students may construct from language and the sorts of difficulties they may encounter in trying to construct meaning from the language that they read and hear in physics.  We are going to suggest that there are some student difficulties that may be recognized primarily as difficulties with language.  Below we present two initial theoretical points that will help the reader understand the role of language in learning and communicating physics.

\subsubsection{Language as a Representation}

We will adopt Jay Lemke's view that the primary activity that students encounter and participate in, in a physics course, is {\it representing} \cite{lemke2004}.  They encounter many different representations of physics ideas: graphs, equations, tables, pictures, diagrams, and words.  These representations of physics ideas are each by themselves incomplete.  It takes an act of assimilating, coordinating, and moving between many different representations in order to create understanding.  Therefore one of the first abilities students have to develop is the ability to represent ideas and physical processes in different ways and move between representations.  Physicists are conscious of the role of equations and graphs in their reasoning.  Less attention, however, has been paid to language as a representation of knowledge and ideas in physics.  Our starting point will be to treat language as a legitimate representation of physical ideas and processes.   Physicists are aware that some student difficulties may be caused by confusing language (see for example, \cite{arons1997,bauman1992,hobson2001,leff1995,mcclain1990,mcdermott1994,romer2001,styer2001,williams1999,zemansky1970}), but only a relatively small amount of research has been done in this area~\cite{lemke1990,touger1991,itzaortiz2003}.  

\subsubsection{Information and Communication}

Reddy~\cite{reddy1993} has suggested that people construct meaning from the words that they hear, based on their prior knowledge and experience.  For example, If you ask someone: ``Are you sad?''  And they respond with a ``1'':  What would  this response mean?  By itself it means nothing, it is simply a signal.  Imagine that you and your partner have a list of possible responses: either 1, 2 or 3.  This is called a  ``repertoire'' of responses.  After the two of you have established a repertoire of responses you need to assign a meaning to them.  Say you two agree before hand that 1 = ``yes,'' 2 = ``no,'' and 3 = ``unable to give a definite answer.''  Now you both have established a shared repertoire of acceptable signals, plus a shared code.  You and your partner have the means to communicate.  This example shows that by itself the signal is meaningless.  A recipient has to construct the meaning using a commonly understood repertoire and a previously shared code (shared a priori between sender and receiver).

From the above discussion, it follows that meaning cannot be directly passed, conveyed or in any way transported from the instructor to the student.  The teacher has to help the student construct meaning by elaborating the code.  Students can then use this code to decode the words that the instructor uses.  For example, when a physicist says ``the electron is in the ground state,''  she means that the electron has a particular energy.  However, if the students do not share the code for the word ``state'' as the energy state, they may construct a spatial interpretation from the same statement.

\subsection{Overview of the paper}

In Section~\ref{framework} we will explain our theoretical framework in the context of the data we gathered.  First, we will describe our language data sources (QM textbooks, interviews with physics professors, and videos of QM students working on QM problems.).  To be able to explain how language works in physics, we found it necessary to introduce the formal categories of analogy, metaphor, grammar and ontology.  We will elaborate how analogy, metaphor, grammar and ontology fit together to describe physicists' language when they ``speak and write physics''.  Finally, we will present two hypotheses about how one can use our theoretical framework to understand how students are interpreting the language that they read and hear in a physics class.

In Section~\ref{qmmetaphors}  we will return to the interviews with physics professors and the QM textbooks and code the language used.  By looking for patterns of usage that can be described and explained by the theoretical framework we have developed, we will show how this framework is applicable for understanding how physicists use language in their reasoning process.

In Section~\ref{student_difficulties} we will describe how we tested the applicability of our theoretical framework for understanding students' reasoning and learning in physics.  We will present two case studies from our video data of physics students working on QM problems.

In Section~\ref{futuredirections} we will explore some future directions that this research on language in physics could proceed.

\section{Theoretical Framework\label{framework}}

\subsection{Introduction}

Our theory was developed from a number of sources of data: (1) Interviews with 5 physics professors. (2) Original QM papers from Born~\cite{born1926} and Schr\"odinger~\cite{schrodinger1982}, as well as an analysis from Goldstein~\cite{goldstein1980} of how Schr\"odinger developed the wave equation. (3) A selection of older and more modern, popular introductory quantum mechanics textbooks~\cite{dicke1960,eisberg1974,french1978,merzbacher1998,griffiths2005,feynman1965}. (4) Two physics student homework study groups.

We began by comparing the way Sch\"odinger and Born wrote about their ideas with the way modern textbooks and physics professors wrote and spoke about the same ideas.  This  led us to define two separate patterns of language used to express ideas in QM:

\begin{enumerate}
\item  The first pattern was language used by the inventors of QM.  They tended to use cautious and figurative language. Ideas were often expressed as comparisons of the form ``X is like Y in certain respects.'' They made analogies explicit and cautioned against overextending or misinterpreting these analogies.
\item The second pattern we observed was language used to  communicate already established knowledge of QM. (Language used by physics professors and modern QM textbooks.) This language was characterized by statements of fact with little if any reference to the original analogies on which the ideas were based. 
\end{enumerate}
These two patterns of language lead us to investigate the role of analogy and metaphor in describing physicists' language.

In our data of students working on their QM homework problems, we focussed our attention on episodes when students stopped calculating, and engaged in an activity that could loosely be described as  {\it sense-making}.  In these episodes it appeared to us as if students understood the physical ideas, but they were confused about the language used to express the physical ideas.  We hypothesized that students were confused by the figurative language that physicists used to describe their ideas. We will justify this claim in Section~\ref{student_difficulties}.

\subsection{Analogical Models Encoded as Metaphors in Physics}

\subsubsection{Metaphors in Physics Language}

{Lakoff and Johnson~\cite{lakoff1980} have hypothesized that human language and the human conceptual system are largely made up of unconscious conceptual metaphors.  We have extended this idea to physics by suggesting that physicists speak and write using conceptual metaphors. For example, physicists talk about ``{\it diffraction} of electrons'' and a ``{\it wave} equation for the electron.''  Both phrases suggest the conceptual metaphor {\sc the electron is a wave}.  Conceptual metaphors are often unconscious metaphors and seldom made explicit. They have become quite literal, losing their figurative origin through their unconscious and frequent use.  For a more complete discussion of what a metaphor is and how it differs from analogy and simile, we refer the interested reader to~\cite{brookes2006b}.  Other excellent discussions of the theoretical status of, and issues surrounding, metaphor and analogy may be found in~\cite{black1962,lakoff2000,atkins2004,bowdle1999,fauconnier2002,glucksberg1990,shen1992,tversky1977,gentner1983,boyd1993}.

\subsubsection{Types of Analogies Encoded as Metaphors}

Researchers have shown that physicists use analogical models to construct new ideas~\cite{hesse1966,nersessian2002}.  These analogical models become, in time,  encoded linguistically as conceptual metaphors~\cite{sutton1978,sutton1993}.  The way physicists talk about already established knowledge is {\it different} than the way they talk about new ideas they are trying to comprehend themselves.

We will take this idea further. From the primary data (textbooks, original papers, and interviews with physics professors), we have identified three types of analogical model that metaphors encode. These can be classified by their origin and function:
\begin{enumerate}
\item {\bf Current analogical models:}  For example, Schr\"odinger based his wave equation on an analogy to wave optics~\cite{schrodinger1982,goldstein1980}.  The corresponding metaphorical system is {\sc the electron is a wave} and is spoken about by modern physicists in terms such as ``electron {\it interference},'' ``electron {\it diffraction},'' ``{\it wave} equation,'' and so on.
\item {\bf Defunct analogical models:}  It is often the case in physics that older models, whose limitations have been experimentally exposed and supplanted by better models, live on in the language of physics.  The caloric theory of heat lives on in phrases which reflect the {\sc heat is a fluid} metaphor.  For example, ``heat {\it flows} from object A to object B.''  Physicists use these metaphorical pictures when they reason.  We will elaborate this point further below.

\item {\bf Descriptive analogies:}  For example an analogy between a physical valley and a potential energy graph.  The metaphor is {\sc potential energy graphs are water wells}.  Examples of how the metaphor is used in language are:  ``potential {\it well},'' ``potential {\it step},'' ``energy {\it level},'' ``{\it ground} state,'' and so on.
\end{enumerate}

We will identify metaphors that encode analogies 1 to 3 by identifying the {\it base} of the analogy~\cite{gentner1983}.  We will use the idea that conceptual metaphors borrow terms from the {\it base} of the analogy and apply these words directly to the {\it target} concept.  For example, if we look at the matter-wave analogy in QM, we can consider that a water wave or an electro-magnetic wave is the prototypical example which will serve as the ``base'' of the analogy.  Thus words such as ``interfere,'' ``polarize,'' ``diffract,'' and ``wave'' are used in the context of  ``an electron.''  Such examples will be identified as instances of the {\sc electron is a wave} metaphor.

\subsubsection{Features and Functions of Metaphors in Physics}

We hypothesize that physicists unconsciously prefer to speak and write in metaphors because metaphors have certain features and functions that are advantageous to them.  The features and functions of these metaphorical systems are listed below with examples from interview data with physics professors.

\underline{\bf Feature 1:} Conceptual metaphors encode analogies.  They encode a more deep and complex piece of knowledge which is the completely elaborated analogy.  That elaboration as an analogical model is, however, tacit amongst the community who use the metaphor and associated model regularly.  

\underline{Function:} Physicists are able to use these metaphorical systems to reason productively about a particular situation or problem.  For example, the {\sc electron is a wave} metaphor can be used productively to explain the Heisenberg uncertainty principle:

\begin{quotation}
$\!\!\!\!\!\!\!\!\!$``I often think of it\ldots in terms of Fourier transforms and the reciprocity between the bandwidth of the channel and the length of the signal pulse that can be detected.'' (Prof A)
\end{quotation}
Note the use of words from the base domain of electromagnetic waves: ``Fourier transforms,'' ``bandwith,'' and ``signal pulse'' in particular.

Even defunct analogies (type 2) represent  productive modes of thought for physicists. There is a class of problems for which it is quite adequate to talk about {\sc heat as a fluid}. For example, when there is no work being done on or by the thermodynamic system, it is satisfactory to think of heat flowing into or out of the system and that the change in temperature of the system is directly proportional to the amount of heat gained or lost.}

\underline{\bf Feature 2:} Metaphorical systems are partial in nature.  This means that more than one metaphorical system is needed to fully understand a physical concept.

\underline{Function:} We observed that physicists switch easily and unconsciously between one system and another depending on the type of question that is asked.  For example, in the following extract, Prof. D switches back and forth between particle and wave metaphor to describe the process of electrons passing through a Young's double slit apparatus.

\begin{quotation}
$\!\!\!\!\!\!\!\!\!$``Of course in any one experiment,\ldots you will not observe\ldots an interference pattern on the screen [wave metaphor] --- if all you do is to scatter one electron [particle metaphor]. The intensities are just too low [wave metaphor]. \ldots you have to have a large number of electrons [particle metaphor], you have to have a beam of electrons [wave metaphor].  And each electron will contribute a little piece of the intensity that you see on that screen [particle metaphor]. What I envisage is\ldots a beam of electrons which can be represented by a plane wave [wave metaphor]\ldots''
\end{quotation}

\underline{\bf Feature 3:} Metaphors involve the use of the verb ``is'' rather than ``is like.''  Metaphors are grammatically {\it identifying relational processes}, i.e., they are grammatically equivalent to statements of category membership.

\underline{Function:}  We hypothesize that metaphor reflects a particular aspect of an expert physicist's thought process.  The use of metaphor itself rather than simile is significant.  Irrespective of deep philosophical discussions about what is ``real,'' it seems apparent that physicists themselves need to assert something stronger than ``like'' --- they need to assert ``is'' in their own reasoning process.    We suggest that this is a fundamental trait of how knowledge is generated and represented in physics.  It is significant because such assertions may often {\it conceal} the vague or partial nature of metaphor itself.  

For example, Prof. D provided the following response to the question: What happens to a single electron when it passes through a Young's double slit apparatus?

\begin{quotation}
$\!\!\!\!\!\!\!\!\!$``\ldots [to] understand that experiment, you've got to forget about the idea that an electron is
a particle.  It is not a particle in that context, it behaves like a wave. {\em So you just think of it as a plane wave} [our emphasis] advancing on the two slits, and the interference between the two\ldots outgoing beams, just using Huygen's principle, leads to the\ldots interference pattern\ldots''
\end{quotation}

Note that comparison, ``it behaves like a wave,'' is followed directly by, ``just think of it {\em as} a plane wave.''

\underline{\bf Feature 4:} The apparatus of language constrains the ways physicists can talk about physical phenomena and therefore constrains the types of models that can be represented in language.
 
\underline{Function:} Descriptive analogies (type 3) encoded as metaphors also represent ways of speaking about/describing physical systems.  This is very important because there is a limit on what can be represented with language.  Such metaphors also give abstract concepts and quantities a grounding in physical reality and physical experience.
 
Consider for example, the modern physicist's view of energy.  Physicists can define energy as a state function yet can physicists speak literally about energy as a state function?  Our hypothesis is that it is simply impossible to come up with grammatical constructions that convey the meaning of energy as a state function.  The very best locutions are ``energy flows into the system,'' or ``process X caused the kinetic energy of the system to increase.''  In both these cases, metaphorically, energy is being spoken of as  {\it matter} and the system as a {\it container} of energy.  (This is suggested particularly by the use of the prepositions ``into'' and ``of.'') It is no coincidence that these two locutions are identical to examples given by Lakoff and Johnson~\cite{lakoff1980}.  The authors describe similar metaphorical patterns in how humans (in English at least) encode physical processes and events as movement of substances into and out of containers.

Physicists are aware of the limitations of their language.  When asked about what is oscillating in a quantum mechanical wave, one professor responded:
\begin{quotation}
$\!\!\!\!\!\!\!\!\!$Prof B: ``The problem is you're trying to shoehorn a phenomenon into ordinary everyday English language, and I think the problem is with the language, not with the phenomenon.  So, if you ask me to explain it in English, I think English has limitations which make it impossible to give a  satisfactory explanation in English.  But, I don't have to understand it in English.  I mean, I think I sort of know what's going on.  At least I have realized the limitations in English and, it doesn't bother me.''
\end{quotation}

\subsection{Ontological Underpinnings\label{ontological_underpinnings}}

\subsubsection{Introduction}
It has been suggested that humans divide the world into ontological categories of {\it matter}, {\it processes} and {\it mental states}~\cite{chi1994}.  In this section we will show that this idea can be applied to models in physics.  The elements of a physical model: the objects or systems of objects, interaction laws, force laws, state laws etc., may be mapped to the ontological categories of matter, processes, and physical states.
In cognitive linguistics, Lakoff and Johnson~\cite{lakoff1980} have shown that systems of conceptual metaphors are based on ontological metaphors.  These ontological metaphors often give abstract concepts an existence as concrete objects or things.  To unite these two views and systematize our linguistic analysis, we hypothesize that ontological metaphors in physics language are realized as grammatical metaphors. Functional grammarians have suggested~\cite{halliday1985} that the elements of a sentence can be divided into {\it participants} (nouns or noun groups), {\it processes} (verbs or verb groups), and {\it circumstances} (generally adverbial or prepositional phrases).  In order to unify the metaphorical and grammatical views, we have suggested~\cite{brookes2005,brookes2006b} that grammatical participants should be mapped to the ontological category of matter, and grammatical processes represent ontological processes.  Ontological physical states also have unique grammatical representations, through the use of grammatical location.

\subsubsection{A Lexical Ontology}

We hypothesize that the concepts in a physical or analogical  model can be arranged into an ontological tree similar to the one proposed by Chi et al.~\cite{chi1994}.   It is necessary to modify Chi et al.'s ontology tree to accommodate one missing category:  namely {\it physical states}.  (See Fig.~\ref{newontology}.)

{\begin{figure*}[htb] \centering
\includegraphics[height=3.5cm]{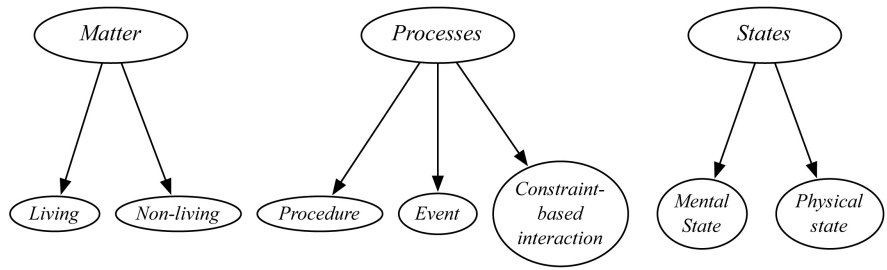}
\caption{A revised ontology tree based on~\cite{chi1994}\label{newontology}}
\end{figure*}\par}

Etkina et al. have suggested that physical models can be broken up into a taxonomy of (1) models of objects, (2) models of interactions between objects, (3) models of systems of objects, and (4) models of processes that the objects/system undergoes~\cite{etkina2006a}.  In addition to their taxonomy, we are going to suggest that there are two classes of physical variables that describe a system or the objects in it.  These are (5) {\it physical properties} of objects (such as mass and charge), and (6) {\it state variables} that describe a configuration of the system (e.g., position, momentum) or {\it state functions} defined over a system configuration (e.g., energy, entropy).

We will now show how Etkina et al.'s model taxonomy can be mapped into the ontology tree shown in  Fig.~\ref{newontology}.  This mapping  is shown in Table~\ref{taxonomy}.

{\begin{table*}[htbp]\centering
\caption{Table illustrating how Etkina's model taxonomy successfully maps into Chi's (modified) ontology\label{taxonomy}}
\begin{tabular}{|p{3cm}||p{1.5cm}|p{1.5cm}|p{1.5cm}|p{4cm}|p{3cm}|}
\hline
{\bf \centering Ontological category} & \multicolumn{2}{p{3cm}|}{\it \centering Matter} & \multicolumn{2}{p{5.5cm}|}{\it \centering Process} & {\it \centering State} \\
\hline
{\bf \centering Ontological sub-category} & \multicolumn{2}{p{3cm}|}{\it \centering Non-living} & {\it \centering Event} & {\it \centering Procedure and Constraint-based Interaction} & {\it \centering Physical State} \\
\hline\hline
{\bf \centering Taxonomy element} & objects & system & interaction laws & causal laws, state laws & state variables, state functions \\
\hline
\end{tabular}
\end{table*}\par}

{\it Physical properties} such as mass and charge should be considered properties of objects classified in the {\it matter} category.

The categorization of concepts in physics into an ontology tree (as shown in Table~\ref{taxonomy}), will be termed a {\it lexical ontology}.  For example, physicists generally agree that energy is a {\it state function}, while heat and work are {\it processes} by which energy is transfered into or out of a system.  Thus a {\it lexical ontology} refers definitions of physics concepts  into {\it matter}, {\it processes}, and {\it states} that physicists would agree with as a community.

\subsubsection{Grammar and Ontology}

Although physicists can agree on the meaning of terms, how do they represent the ontology of physics concepts with language?

We suggest that every physical model described in language has an ontology and that this ontology is encoded in the {\it grammar} of the sentence.  This grammatical ontology can be either literal or figurative (metaphorical).  If the {\it lexical ontology} matches the {\it grammatical ontology} then the sentence is literal. If the {\it lexical ontology} does not match the {\it grammatical ontology} of the same term in a given sentence, then a grammatical metaphor is present.  We suggest that these metaphors may be consistently identified by using the grammatical/ontological analysis elaborated below.  For an introduction to the methods of functional grammar, we refer the reader to~\cite{halliday1985}.

Consider for example, ``John [{\it agent}] kicked [{\it process}] the ball [{\it medium}].''  Here ``John'' and ``the ball'' are grammatical participants, functioning grammatically as objects or {\it matter}.  We also recognize that ``John'' and ``the ball'' are naturally defined as {\it matter} in some sense.  Thus the grammatical ontology and {\it lexical ontology} match.  There is nothing metaphorical in this sentence.  Consider now for example,  ``heat [{\it medium}] flows [{\it process}] from the environment to the gas.''  In this sentence a physicist would recognize heat to define a {\it process} of movement of energy into the system ({\it lexical ontology}).  But grammatically ``heat'' is functioning as a {\it participant}, namely heat is the {\it matter} that is flowing.  In this case the grammatical function of the term ``heat'' and the {\it lexical ontology} of ``heat'' contradict each other.  The sentence is therefore metaphorical.

We are going to propose the following mapping from grammar to the ontology tree shown in Fig.~\ref{newontology}: Grammatical {\it participants} should be mapped into the ontological category of {\it matter}.  {\it Participants} can immediately be separated into {\it living} and {\it non-living} ontological subcategories:  {\it Beneficiary}, {\it agent}, and {\it medium} (as it participates in an {\it action process}, such as ``a force [{\it medium}] acts [{\it process}]''), can all be thought of as {\it living} entities.  {\it Range} and {\it medium} (as it participates passively in an {\it event process} such as ``heat [{\it medium}] flows [{\it process}]'') can be thought of as {\it non-living} entities.

Certain parts of {\it circumstantial} elements can also be mapped to the {\it matter} category.  In the example, ``\ldots the incident particles [{\it medium}] will be\ldots partially transmitted [{\it process}] through the potential-well region [{\it location}].''   ``the potential-well region'' could be classified as  {\it non-living matter}.  However {\it location} also functions grammatically to make ontological {\it physical states} as in ``A particle [{\it medium}] is [{\it relational process}] at coordinates (1,1,1) [{\it location}].''  This will be discussed further below.

An important type of grammatical {\it process} in the discourse of physics is the {\it relational process}.  {\it Relational processes} are processes of being in that they almost always include some form of the verb ``to be.''  {\it Relational processes} have two modes: {\it identifying} and {\it attributional}.    The {\it identifying} mode is a reflexive relationship. For example, ``the neutrino is the lightest known particle.''  It makes sense to say ``the lightest known particle is the neutrino.''  The {\it attributional} mode denotes category membership and is not reflexive.  For example: ``An electron is a lepton.''

We hypothesize that physical states (as expressed in physicists' language) are commonly comprised of {\it identifying relational processes} where the second identifier is missing and replaced by a grammatical {\it circumstance} of {\it location}.  Typical examples are: ``The electron is in the ground state,'' ``the particle is at such and such coordinates.''   Such sentences very often involve a grammatical metaphor.  Ontologically {\it location} is mapped to some sort of physical object or {\it matter}, this often conflicts with the lexical ontology.  These grammatical metaphors correspond directly to the ontological metaphors of Lakoff and Johnson in choice of preposition: ``in'' implies container, ``at'' implies point location in either time or space, ``on'' implies surface.  We believe that it is also no coincidence that these statements have a grammatical structure identical to those of mental states.  For example, in English we say, ``I am {\it in} love,'' or ``I am in trouble,'' or ``I am in a state of confusion'' etc.  It seems to us that physicists have borrowed this metaphor wholesale and blended it with the notion of a physical state, to create a way of speaking about physical states.

Ontological {\it processes} that describe the {\it behavior} of a physical system, are realized in speech by grammatical {\it material processes}.  {\it Relational processes} realize either {\it physical states} as shown above, or denote some component of the model in the sense of category membership.   In grammar there are two types of {\it material process}: {\it action} and {\it event}.  These two types of {\it process} can be used to distinguish between {\it living} and {\it non-living} matter.

The entire mapping from grammar to ontological category is summarized in Table~\ref{grammar_ontology_mapping} below.

\begin{table}[htbp]
\caption{Summary of the mapping between grammar and ontological category \label{grammar_ontology_mapping}}
\begin{ruledtabular}
\begin{tabular}{p{3.5cm}cp{3.5cm}}
\parbox{3.5 cm}{\bf \centering Grammatical function} & & \parbox{3.5 cm}{\bf \centering Ontological category} \\
{\it (If X functions grammatically as\ldots)} & $\longrightarrow$ & {\it (\ldots classify X ontologically as\ldots)}  \\
\hline
{\it Agent} & $\longrightarrow$ & {\it Matter:living}  \\
{\it Beneficiary} & $\longrightarrow$ & {\it Matter:living}  \\
{\it Medium} (action process) & $\longrightarrow$ & {\it Matter:living}  \\
{\it Medium} (event process)  & $\longrightarrow$ & {\it Matter:non-living}  \\
{\it Role} & $\longrightarrow$ & {\it Matter:non-living}  \\
Objects in {\it Location} & $\longrightarrow$ & {\it Matter:non-living} \\
{\it Process} & $\longrightarrow$ & {\it Process} \\
{\it Manner} & $\longrightarrow$ & {\it Process} \\
\end{tabular}
\end{ruledtabular}
\end{table}

\subsection{Summary}
The theoretical framework is summarized in Fig.~\ref{languagesummary} below.
{\begin{figure}[htbp] \centering
\includegraphics[width = 6cm]{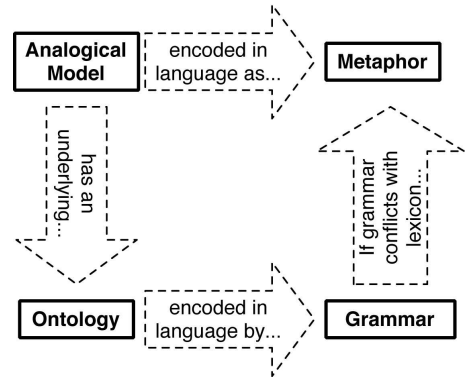}
\caption{Summary of the role of analogy, metaphor, ontology and grammar\label{languagesummary}}
\end{figure}\par}
Consider, for example, the caloric theory of heat.  This theory  of thermodynamics began as an analogy to a weightless fluid in the late eighteenth century.  Over time the elements of this theory became encoded in the language of physics as a conceptual metaphor.  For example, physicists today still say ``heat {\it flows} from object A to object B,'' and talk about the ``heat {\it capacity}'' of an object.  Phrases and sentences such as these are evidence of the conceptual metaphor {\sc heat is a fluid} in physicists' language.  For physicists, speaking about heat is a fluid is a productive mode of reasoning as long as there is no work being done on or by the thermodynamic system.  The applicability and limitations of talking about heat as a fluid are communally well understood.  The analogy between heat and a fluid has an underlying ontology of {\it matter} (the heat fluid), {\it processes} (the movement of heat from one object to another), and {\it states} (the amount of heat {\it in} an object --- indicated by the object's temperature).  This ontology is encoded in the grammar of each sentence used to speak or write about the thermodynamic system.  In the example, ``heat flows from object A to object B,'' heat is a grammatical participant, while the grammatical process is ``flows''.  Object A and B are parts of grammatical location.  Implicitly, the amount of heat in object A or object B indicates the current state of the system.  In the modern thermodynamic model, the ontological {\it matter} is the atoms or molecules in the system, the {\it processes} that the system undergoes are heating and work (processes of energy transfer), and the {\it state} of the system is represented by the energy or entropy of the system for a given configuration of the molecules.  Note how the modern ontology is in direct conflict with the caloric ontology of thermodynamics.  Speaking about heat as matter is therefore a grammatical metaphor.  It is grammatical metaphors like this that underpin the conceptual metaphor {\sc heat is a fluid}.

\subsection{Student Difficulties, Student Learning\label{student_reasoning}}

The central question of our paper is: What is the interplay between the linguistic representations that physicists use and students' learning and students' difficulties?  We will narrow this down to two hypotheses regarding the role of language and learning in physics.  These are elaborated in Sections~\ref{h1} and~\ref{ontological_confusion} below.

\subsubsection{Student Difficulties Interpreting Metaphors\label{h1}}

Students struggle to see the applicability and limitations of analogies that they encounter.  We suggest the same applies to metaphorical language that they hear and read.  To comprehend a metaphor people construct an ad hoc category~\cite{glucksberg1990,shen1992}.  This means that a statement of the form ``X is Y'' has to be interpreted through the formation of a {\it new shared category} (an {\it ad hoc category}) of which Y is a prototypical member.  For example, to comprehend a metaphor such as {\sc the electron is a smeared paste}, the reader has to come up with an ad hoc category shared by both entities.  A physicist who understands the quantum mechanical behavior of an electron, might suggest an ad hoc category of ``things that don't have a well-defined location.''   There is no guarantee that a student will come up with the same classification.  We hypothesize that students often come up with an ad hoc category that is inappropriate to a given situation.  This inappropriate categorization is at the heart of their difficulties. These difficulties may manifest themselves as ``misconceptions'' or student difficulties.   We predict that students will overextend and misapply key aspects of metaphorical systems in physics.  Instances where metaphors are overextended or taken too literally will be connected with their faulty reasoning.

In order to test these ideas it is first necessary to identify if there are really coherent systems of conceptual metaphors in the way physicists speak and write.  In Section~\ref{qmmetaphors} we will show some of the interview data with physics professors that lead us to see that this view of language was really applicable to the discourse of physics.  In Section~\ref{student_difficulties} we will consider examples of student difficulties in QM that we can explain as examples of metaphorical overextension.

\subsubsection{Students' Ontological Confusion\label{ontological_confusion}}

Previously, Chi et al. have shown that many student ``misconceptions'' are based on students' incorrect ontological classification of physics concepts~\cite{chi1994,reiner2000,slotta1995}.  For example, physicists classify heat is a {\it process}, but students reason with it as if it were {\it matter}.

We want to propose an extension of this idea.  From our data it appears that physicists reason about physics by co-ordinating multiple descriptions of a particular phenomenon. These descriptions may possess different or conflicting ontological properties.  For example, there are times when physicists talk about QM phenomena in terms of waves (a {\it process} description) whereas there are other times when  physicists prefer to talk about a QM phenomenon in terms of particles (a {\it matter} description).  Physicists are good at co-ordinating these different and sometimes conflicting descriptions.   Physicists understand when a wave or a particle description work and how use them appropriately in their reasoning.

Students learn these descriptions by listening to and reading what their teachers say and write.  Our second hypothesis is that students are failing to co-ordinate appropriately the many different descriptions that they learn from physicists' language.  For example, physicists often describe the ``potential energy graph'' in QM in terms of physical objects (well, barrier, etc.), endowing the graph with the properties of a physical object.  Students, hearing this language, also learn to think of the graph as a physical object.  However, students are often unaware of when this picture is appropriate or inappropriate.  Thus students may attach inappropriate ontological properties to the idea of the graph as a physical object.  We hypothesize that this process leads to patterns of student reasoning that researchers sometimes interpret as ``misconceptions''.

In Section~\ref{student_difficulties} we will test this hypothesis by analyzing an example of students solving a QM problem and consider several studies from the PER literature.

\section{Metaphors in Quantum Mechanics\label{qmmetaphors}}

We will trace two metaphorical systems in QM from their origins as analogies through to modern language that physicists use to speak and write about their ideas.   These two systems are (1) the {\sc potential well} metaphor, and (2) the {\sc Bohmian} metaphor.

Both grammatical and metaphorical analyses will serve together to illustrate a number of claims made in Section~\ref{framework}.  The claims are: (1) Coherent systems of metaphors exist in physicists' language. (2) Systems of metaphors encode historical analogies. (3) The language encodes a representation or representations of a physical model that has an underlying ontology of {\it matter}, {\it processes} and {\it states}. (4) Physicists use these linguistic representations to reason productively about certain phenomena.

The data for the linguistic analysis will come from two sources.  The first is the interview study with physics professors referred to in Section~\ref{framework}.  These professors were all native English speakers   We  asked them to describe and explain various ideas in QM such as the Heisenberg uncertainty principle, or how they would respond to a student who asked, ``what is oscillating in a QM wave?''  The interview study consisted of five subjects.  The full set of interview questions may be obtained by request from the authors.  The second source of data is a selection of QM textbooks~\cite{dicke1960,eisberg1974,french1978,merzbacher1998,griffiths2005,feynman1965}.

\subsection{The {\sc Potential Well} Metaphor}

\subsubsection{Original Descriptive Analogy}

\begin{quotation}
$\!\!\!\!\!\!\!\!\!$``Because of the Pauli exclusion principle, the electrons must be spread over the available states; but they settle down to the states of lowest energy, so that as more electrons are added, the energy levels in the band fill up like a bucket fills with water.''~\cite{peierls1985}
\end{quotation}
In this example Peierls makes the analogy explicit.  The way he uses it shows that this analogy has a descriptive role

\subsubsection{Analysis of Modern Language\label{potential_well_analysis}}

\underline{\bf Grammatical and ontological analysis:} When physicists speak of ``potential {\it well}'' and ``energy level,''  they give energy an existence as water.  When physicists speak about quantum particles ``{\it leaking} through a barrier,'' they give the quantum particles an existence as water.  When physicists speak of a ``potential {\it well},'' ``potential {\it step},'' ``potential {\it barrier},'' ``confinement,'' ``trap'' etc\ldots they give the potential energy graph an existence as a physical object.  This ontology is encoded in the grammar.  Samples of textbook writing and physicists' talk from interview data and accompanying grammatical analysis are provided in Table~\ref{potential_well_grammar}.

\begin{table*}[htbp]
\caption{Samples of physicists' speech and writing for grammatical analysis.\label{potential_well_grammar}}
\begin{ruledtabular}
\begin{tabular}{p{8.5cm}p{8.5cm}}
{\bf Sample of physicist's speech or writing} & {\bf Simplified exerpt with analysis} \\
\hline
 ``In both cases, a classical particle of total energy $E$\ldots moves back and forth between the boundaries.''~\cite{french1978} & a classical particle [{\it medium}] moves back and forth [{\it process:event}] between the boundaries [{\it circumstance:location}]. \\
& \\
``\ldots when you have a confined system, \ldots [the width of the box is] going to set the scale for what \ldots the magnitude of the energy is, so as you confine it [the particle] more and more, your zero point energy is going to go up and up.'' - Prof. A, interview study & \ldots your zero point energy [{\it medium}] is going to go up and up [{\it process:event}]. \\
& \\
``\ldots it has been seen that potential barriers can reflect particles that have sufficient energy to ensure transmission classically.''~\cite{dicke1960} & \ldots potential barriers [{\it agent}] can reflect [{\it process:event}] particles [{\it medium}]\ldots \\
& \\
``This [wave] packet would move classically, being reflected at the wall\ldots''~\cite{merzbacher1998} & This wave packet [{\it medium}]  is reflected [{\it process:event}] at the wall [{\it circumstance:location}]. \\
& \\
``The $\alpha$-particle then `tunnels through' the barrier\ldots''~\cite{dicke1960} & The $\alpha$-particle [{\it medium}] then `tunnels through' [{\it process}] the barrier\ldots [{\it range}] \\
& \\
``\ldots they [$\alpha$ particles] start out with the energy $E$ {\it inside} the nucleus and {\it `leak' through} the potential {\it barrier}.''\cite{feynman1965} & $\alpha$ particles [{\it medium}] leak through [{\it process:event}] the potential barrier [{\it range}]. \\
& \\
``\ldots the phenomenon of tunneling\ldots allows the particle to `leak' through any finite potential barrier\ldots''~\cite{griffiths2005} & The particle [{\it medium}] leaks through  [{\it process}]  any finite potential barrier [{\it range}]. \\

\end{tabular}
\end{ruledtabular}
\end{table*}

From the data we have studied, this selection of talk and writing of physicists (Table~\ref{potential_well_grammar}) is representative of the type of language associated with the {\sc potential well} metaphor.  One can see a clear pattern of grammar that can be mapped to the ontological categories of {\it matter} and {\it processes}.  This is shown in Table~\ref{potential_well_ontology} below.

\begin{table}[htbp]
\caption{Ontology of the {\sc potential well} metaphor\label{potential_well_ontology}}
\begin{ruledtabular}
\begin{tabular}{p{5.5cm}p{2.5cm}}
\parbox{6cm}{\bf \centering Matter} & \parbox{2cm}{\bf \centering Process} \\
\hline
QM/classical particles, wave packet, energy, energy walls, energy barrier, potential barrier, barrier & moves, reflect(s), tunnels through, leaks through \\
\end{tabular}
\end{ruledtabular}
\end{table}

The grammatical analysis can tell us more than what the objects and processes are in the metaphorical model.  It shows us that the potential well metaphor consists of two objects, the particle/wave function, and the potential energy graph, that function as {\it separate} grammatical participants.  They interact with each other via a number of possible processes such as ``tunnel through,'' ``reflects,'' and so on.  Thus the common grammatical structure of the {\sc potential well} metaphor contradicts the conventional view of the potential energy as a {\it property of} the particle or the system.

\underline{\bf Metaphorical analysis} The ontology encoded in the grammar describes the basic objects and processes of the physical model.  To understand more subtle properties of these objects, and their interactions, we need to apply a metaphorical analysis.   For this, we need to identify the base of domains of various analogs that go into making up the {\sc potential well} metaphor.

In this section we will analyze an additional sample of clauses and sentences from a selection of popular introductory quantum mechanics textbooks~\cite{dicke1960,eisberg1974,french1978,merzbacher1998,griffiths2005}.  We will identify each metaphor that makes up the metaphorical system, and present sample examples of its use by physicists. {Additional examples may be found in~\cite{brookes2006b}.) Where necessary, we will identify the analogical {\it base} from where the words have been ``borrowed'' to create the metaphor.

\begin{itemize}

\item \underline{\bf Metaphor:} {\sc The potential energy graph is a physical object or physical/geographical feature.}

\underline{\bf Examples:}``The perfectly rigid {\bf box}, represented by a rectangular potential {\bf well} with infinitely high walls, is an ideally simple vehicle for introducing the mathematics of quantum systems.''~\cite{french1978}

``Scattering from a `{\bf cliff}'.''~\cite{griffiths2005}

``\ldots even for a total energy of the particle less than the maximum height of the potential {\bf hill}\ldots''~\cite{dicke1960}

``What are the classical wave analogs for particle reflection at a potential {\bf down-step} and a potential {\bf up-step}?''~\cite{french1978}

\underline{\bf Base:} The words ``box,'' ``well,'' ``hole,'' ``cliff,'' and ``hill'' are borrowed from the category of  physical objects or physical/geographical features.

\item \underline{\bf Metaphor:} The previous metaphor {\sc the potential energy graph is a physical object or physical/geographical feature} entails another metaphor: The ``walls'' of the well or barrier correspond to a physical height above the ground.  In other words, {\sc energy is a vertical spatial dimension} in the Earth's gravitational field.  {\sc The potential energy graph is a physical object or physical/geographical feature} metaphor builds on this spatial metaphor.

\underline{\bf Examples:}``It is instructive to consider the effect on the eigenfunctions of letting the walls of the square well become very {\bf high}\ldots''~\cite{eisberg1974}.

Prof A: ``\ldots your zero point energy is going to go {\bf up} and {\bf up}.''

``\ldots  $\displaystyle \psi_1$, which carries the {\bf lowest} energy, is called the {\bf ground} state\ldots''~\cite{griffiths2005}

\underline{\bf Base:} The words ``high,'' ``up,'' ``lowest,'' and ``ground'' all suggest an analogy between the vertical axis of the potential energy graph and a vertical spatial dimension on the Earth's surface.

\item \underline{\bf Metaphor:} {\sc The potential energy graph is a container.}  The potential energy graph ``contains'' or ``traps'' either the wave function, the particle or the energy of the particle.

\underline{\bf Examples:} ``The exponential decrease of the wave function {\bf outside} the square {\bf well} for the second energy state is less rapid than is the corresponding decrease for the lowest energy state as indicated\ldots''~\cite{french1978}

``{\bf Inside} the {\bf well} where $\displaystyle V(x) = -V_0$\ldots''~\cite{griffiths2005}

``\ldots {\bf bound} [energy] states {\bf in} the\ldots well''~\cite{french1978}

\underline{\bf Base:} Words such as  ``well,'' ``confined,'' ``bottle,'' and ``bound'' all suggest an analogy to some sort of container.  Physicists also make a distinction between ``in/inside'' and ``outside'' the well:  Such adverbial phrases also indicate the presence of the container metaphor.

\end{itemize}
Various other elaborated metaphors are built on this basic set.  Examples of their usage may be found in~\cite{brookes2006b}.

\begin{itemize}

\item {\sc The potential energy graph is a barrier.}

 \item {\sc The potential energy graph is a hard container/barrier} or {\sc the potential energy graph is a semi-hard container/barrier}

\item The particle, the wave packet, and the energy are all given an ontological status of {\it matter}.  More specifically:

\begin{itemize}
 \item {\sc QM particles are hard objects.}

 \item {\sc The wave packet is a soft or breakable object}

 \item {\sc QM particles are a fluid}

 \item {\sc The energy is a fluid}

 \end{itemize}

\end{itemize}

See Table~\ref{potentialwellmapping} for a summary of the metaphorical mapping from the domain of physical/geographical features to the domain of quantum systems that involve an interaction between two or more objects.

\begin{table}[htbp] \centering
\caption{Summary of the metaphorical mapping between the {\it base} domain of physical/geographical features and the {\it target} domain of interacting QM systems\label{potentialwellmapping}}
\begin{ruledtabular}
\begin{tabular}{p{3.7cm}cp{4cm}}
\parbox{3.7cm}{\it \centering Base domain}  & & \parbox{4cm}{\it \centering Target domain } \\
{\sc Physical/geographical features} & & {\sc Interacting QM systems} \\
\hline
Physical or geographical features & $\rightarrow$ & Potential energy graph \\
Vertical height of physical/geographical feature & $\rightarrow$ & Magnitude of energy at a point/region on the potential energy graph. \\
Hardness or softness of a wall & $\rightarrow$ & ``Height'' of the potential energy graph \\
Container with top face open & $\rightarrow$ & ``Trapping'' of QM particles, ``bound'' states \\
Billiard ball & $\rightarrow$ & QM particle in some circumstances \\
Soft or breakable objects & $\rightarrow$ & QM wave function or wave packet \\
Fluid & $\rightarrow$ & QM particle in some circumstances, or the energy of the particle/system \\
Ball bounding off a wall & $\rightarrow$ & Reflection of QM particle \\
Tunneling/penetration & $\rightarrow$ & Process by which a QM particle ``passes through'' a seemingly solid ``barrier'' \\
Leaking & $\rightarrow$ & Process by which a QM particle ``escapes'' from a QM ``container'' \\
\end{tabular}
\end{ruledtabular}
\end{table}

\subsubsection{Productive Modes}

How do physicists piece together the grammar/ontology of the {\sc potential well} metaphor? How do they use the associated imagery to reason productively about quantum systems?
From the discourse of professors and textbooks we have identified the presence of  productive modes for the {\sc potential well} metaphor.  We present five examples below:

 \underline{\bf Squeezing:} Squeezing the walls of the well forces the water upwards, thereby raising and spacing out the ``energy levels.''

\underline{Example:} Prof. A: ``\ldots when you have a confined system, \ldots [the width of the well is] going to set the scale for what \ldots the magnitude of the energy is, {\it so as you confine it more and more, your zero point energy is going to go up and up.}''

\underline{\bf Stacking:} Matter takes up space.  Filling up the well/bucket can be used to understand the behavior of fermions.

\underline{Example:}  We already observed Peierls make this analogy explicit\cite{peierls1985}.  The following is an example of Prof. A. using it: ``if you have fermions then\ldots you have to keep stacking the fermions into levels which get more and more elevated in energy\ldots''

\underline{\bf Tunneling/leaking:} The  potential energy graph behaves as a physical container or barrier, preventing the escape of the particle.  This  leads to the ideas of ``tunneling'' or ``leaking.''  When reasoning productively, physicists recognize that a higher or wider barrier means less probability of tunneling.

\underline{Examples:} Feynman et al. write: ``\ldots they [$\alpha$-particles] start out with the energy $E$ {\it inside} the nucleus and {\it `leak' through} the potential {\it barrier}.''\cite{feynman1965}

Griffiths writes: ``If the barrier is very high and/or very wide (which is to say, if the probability of tunneling is very small), then the coefficient of the exponentially increasing term ($C$) must be small\ldots''~\cite{griffiths2005}.

\underline {\bf Reflecting/scattering:} The wave function or particle is reflected by or scatters off a hard barrier. 

\underline{Example:}  ``For this reason, the rectangular potential barrier simulates, albeit schematically, the scattering of a free particle from any potential.''~\cite{merzbacher1998}

\underline{\bf A way of speaking:} It is difficult to come up with realistic physical systems of quantum mechanical particles without resorting to lengthly descriptions.  (See~\cite{french1978} for examples.)  By separating the QM particle from its potential energy graph, physicists are able to talk easily about the particle interacting with an external object (its own potential energy graph).

\underline{Example:} During one of the interviews we asked a professor to describe the process of trapping and cooling atoms to absolute zero.

\begin{quotation}
$\!\!\!\!\!\!\!\!\!$DTB:  ``Are the atoms going to jump out, are you not going to be able to trap them?''

$\!\!\!\!\!\!\!\!\!$Prof. E: ``No, of course not, you'd just go down to the lowest eigenstate.  I mean, I don't know how they were trapped in the first place, but suppose you had them {\it in a square well} for example.'' 
\end{quotation}

\subsubsection{Summary}
We have tried to illustrate how grammar and metaphor work together to encode the features of a particular descriptive model.  Each aspect is necessary and the grammatical and metaphorical analysis together  serve to illuminate features that each individual analysis cannot do on its own.

Fig.~\ref{graphicalsummary} presents a visual summary of how the {\sc potential well} metaphor is structured.  The {\sc potential well} metaphor is an example of a metaphorical system made up of three ontological metaphors, two of which are encoded in the grammar ({\sc the potential energy graph is a physical object or physical/geographical feature}, and {\sc the particle/wave function/energy is a physical object or matter}), and one which can only be identified by looking at the imagery ({\sc energy is a vertical spatial dimension}).  Other metaphors such as {\sc the potential energy graph is a container} or {\sc the potential energy graph is a hard barrier} build on and elaborate this ontology.  At the sentence level, we can see how productive modes of reasoning are formed by introducing grammatical processes through which the particle or wave function interacts with its own potential energy graph.  These productive modes are squeezing, stacking tunneling/leaking, and reflecting/scattering.  Physicists also use the {\sc potential energy graph is a physical object} metaphor as a substitute term (or {\it metonym}) for the actual physical QM system.}

{\begin{figure*}[htbp] \centering
\includegraphics[height = 10cm]{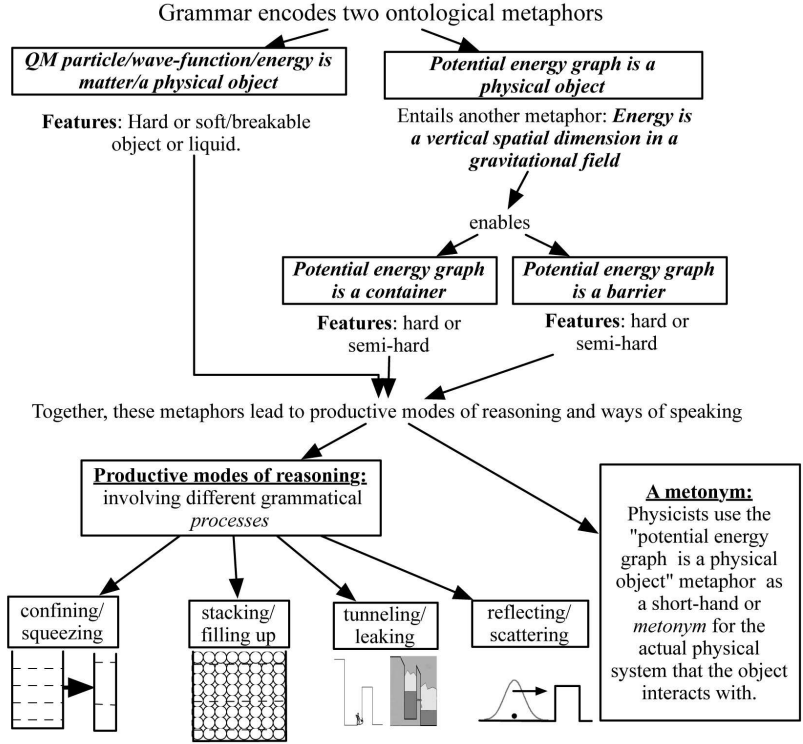}
\caption{Summary of the metaphorical system and its usage by physicists.\label{graphicalsummary}}
\end{figure*}\par}

\subsection{The {\sc Bohmian} Metaphor}

\subsubsection{Introduction}

The {\sc potential well} metaphor, as a linguistic representation, has many of the characteristics of a physical model as described by Etkina et al.~\cite{etkina2006a}.   The language describes objects with properties, and processes by which those objects interact with each other.  In contrast, the {\sc Bohmian} metaphor has almost none of those characteristics.  It seems to exist in the language of physics solely as a way of speaking.  It is an interesting case because it is easy to identify the metaphor, but not the original analogy.  Therefore, in this section, we will present the linguistic analysis first and the study of the analogy on which it is based, second. Although we have called it the ``{\sc Bohmian}'' metaphor in honor of David Bohm, who advocated the Bohmian interpretation of QM, the entry into the language of physics can be traced back much earlier than this.  

\subsubsection{Modern Language}

The {\sc Bohmian} metaphor is identified in language by words and phrases that suggest that the wave function or quantum state is a {\it container} that contains the quantum mechanical particle.  There are only two metaphors that make up the {\sc Bohmian} metaphor:

\begin{itemize}

\item \underline{\bf Metaphor:} {\sc The wave function/quantum state is a container}.

\underline{\bf Examples:} Noun groups such as ``wave {\bf packet}'' or ``{\bf envelope} function'' indicate an analogy to a container.

\item \underline{\bf Metaphor:} {\sc The QM particle is a physical object contained inside the wave function/quantum state}.

\underline{\bf Examples:} This is suggested by prepositional phrases such as ``{\bf in} the ground state'' in sentences such as ``The electron is {\bf in} the ground state.''
\end{itemize}

\underline{\bf Connection to grammar:} In the Bohmian metaphor the wave function or quantum state is conceived of as a {\it container} that has a particle as a separate entity inside it.  The language is based on two sources.  The first source is an analogy to Einstein's ghost field idea (see Section~\ref{einanalogy} below) but the second source is language itself.  Cognitive linguists hypothesize that mental states are spoken about in language metaphorically as containers~\cite{lakoff1980}.   For example, if one is depressed one can say, ``I am [{\it relational process}] in a state of depression [{\it location}].''  Such statements seem to all have the same grammatical structure, namely a {\it relational process} followed by circumstance of {\it location}.  It seems as if ontological {\it physical states} are expressed by an identical grammatical structure:  such as, ``the electron is [{\it relational process}] in the ground state [{\it location}].''   It seems as if physicists have unconsciously borrowed this grammar that expresses {\it mental states} in every day experience and used it to express physical states in physics.  As mentioned in Section~\ref{framework}, the metalingual apparatus that we have for realizing physical states in language appears to be extremely limited. The {\sc states are locations} metaphor, supported by this unique grammatical structure,  is one of these limited means of expression. A statement about the physical location of an object within another object would be classified in the ontological category of {\it matter} if taken literally.  Metaphorically a statement such as ``the electron is in the ground state,'' is a statement about the energy of a quantum system, and physicists recognize energy as a state function.  There is a clear ontological conflict between the literal interpretation of the statement and the meaning that is intended.  This leads us to hypothesize that such statements will cause students confusion and may lead to difficulties.

\subsubsection{The Original Analogy\label{einanalogy}}

In the case of the {\sc potential well} metaphor, the analogy on which it is based, is  relatively well understood.  The {\sc Bohmian} metaphor is easy to identify, but the original analogy is not well known.  If our framework is correct and language is built on analogy then an original analogy should exist in the mainstream QM literature.  We started searching the original QM papers in the hope that we would find some explicit reference to the idea that the wave function could contain the particle inside it.  Remarkably, we found such a reference in a paper by Max Born, published in 1926~\cite{born1926}.

\begin{quotation}
$\!\!\!\!\!\!\!\!\!$``Neither of these two views seem satisfactory to me.  [Heisenberg's interpretation of the wave function and the Schr\"odinger/deBroglie interpretation of the wave function]  I would like to attempt here a third interpretation and test its applicability to collision processes.  I thereby pin my hopes on a comment of Einstein's regarding the relationship between the wave field and light quanta.  He says roughly that the waves may only be seen as guiding [showing] the way for corpuscular light quanta, and he spoke in the same sense of a ``ghost field.''  This determines the probability that one light quantum, which is the carrier of energy and momentum, chooses a particular [definite] path.  The field itself, however, does not have energy or momentum.''~\cite{born1926} [Translation by D.T.B.]
\end{quotation}

There are several remarkable features about this passage from Born:
\begin{itemize}
\item  Firstly, it lays out the Bohmian interpretation of quantum mechanics twenty-five years or more before Bohm proposed the same idea, and one year before deBroglie's attempt at a ``pilot wave'' theory.   
\item Secondly, when Born says ``Neither of these two views seem satisfactory to me,'' he is referring to (1) the Heisenberg interpretation of QM which Born describes as ``an exact description of the processes in space and time are principally impossible,'' and (2) the Schr\"odinger/deBroglie interpretation which Born summarizes: ``He tries to construct wave groups which have relatively small dimensions in all directions and should, as it seems, directly represent moving corpuscles.''  Born is cautioning against overly literal interpretations of (1) an analogy to a classical particle (Heisenberg's approach), or (2) an analogy to a physical wave (Schr\"odinger's approach).   Born suggests that both views lead to untenable positions in the physical interpretation of QM and introduces a third model which is essentially a hybrid of the wave and particle analogies.  Born makes, an analogy to Einstein's interpretation of light waves and light quanta and applies it to particles with non-zero mass.

Born's mode of reasoning appears to be metaphorical as well as analogical.  He makes an analogy to Einstein's view of the electromagnetic field as a ghost field, but he does not suggest that the wave function is ``like a guiding field.''  Rather, he expresses Einstein's idea directly as ``\ldots the waves may only {\bf be seen as} guiding the way for corpuscular light quanta\ldots'' [our emphasis].    For Born to interpret the wave function as a probability distribution, he felt it necessary to blend together a wave picture and a particle picture with {\it real} particles who have {\it definite} trajectories determined probabilistically by the wave function.  Lakoff and N\'u\~nez refer to such a mental construct as a {\it metaphorical blend}~\cite{lakoff2000} after the {\it conceptual blend} of Fauconnier and Turner~\cite{fauconnier2002}.

\item Thirdly, Born is aware  of the limitations of the metaphorical picture he has introduced.  In blending a wave and particle picture into  a model that looks and feels like a statistical ensemble,   Born cautions about taking this ``Bohmian'' picture too literally when he writes: ``However, the proposed theory is not in accordance with the consequences of the causal determinism of single events.''~\cite{born1926}

\end{itemize}

\subsubsection{Productive Modes}

One of the difficulties with QM is the question of how to speak about quantum processes meaningfully.  We suggest that the {\sc Bohmian} metaphor permits a partial solution to this problem.  Although Born's suggestion (intepreting the wave function as a pilot wave) never made it to the mainstream of physics, the associated language is now ubiquitous and used productively by physicists as we will show in the following example:

\begin{quotation}
$\!\!\!\!\!\!\!\!\!$D.T.B.: ``\ldots if you wanted to think about how an electron propagates\ldots It wouldn't be sensible to talk about it as a wave, you would think more as a particle?'' 

$\!\!\!\!\!\!\!\!\!$Prof C: ``\ldots you can think of it as a plane wave.   Yeah, \ldots {\bf in} an {\bf envelope function} which makes it into a wave {\bf packet}.''
\end{quotation}
More examples may be found in~\cite{brookes2006b}.

\section{Student Difficulties\label{student_difficulties}}

\subsection{The {\sc Potential Well} Metaphor}

A group of four junior students in their first QM course, were video taped while working on their QM homework problems.  All students were native English speakers. The discussion we present is centered around a problem  from French and Taylor\cite{french1978}.  The question was: ``What are the classical {\it wave} analogs for particle reflection at a potential down-step and a potential up-step?''
Notice here the {\sc potential well} metaphorical system serving a specific function: namely, it describes the shape of the potential energy graph (``potential down-step'').

\begin{quotation} 
$\!\!\!\!\!\!\!\!\!$S1: Well, there wouldn't be reflection in particle physics on a down-step right?  Or even, I don't think even on an up-step\ldots \newline
S3: No, there's reflection on an up-step, total reflection.\newline
S1: Not classical though, right?\newline
S2: Not if its less than the energy though.\newline
S1: It just slows it down.
\end{quotation}

In this opening exchange we can observe S1 talking at cross purposes with S2 and S3.  S2 and S3 seem to be imagining a classical particle approaching the step and bouncing back (later dialogue show that they do not really shift from this literal view of the situation), while S1 seems to be thinking of a wave approaching with energy greater than the energy of the step.  As we see later, S1 is reasoning from picture of a surface water wave passing over a step in a river or sea bed.

\begin{quotation}
$\!\!\!\!\!\!\!\!\!$S1: Not quite sure what the wave analogs would be.  If I had to guess I'd say it would be like sound, like those things that male cheerleaders have, like big cones.\newline
S4: Megaphones?\newline
S1: Yeah. 'Cause I think, you know,\ldots basically a step up or step down in resistance.  But I am not quite sure what we are supposed to say about that.
\end{quotation}

This is the first example of an analog from S1.  It is interesting that S1 sees the key as a change in resistance  (at the end of the first exchange S1 says ``It just slows it down''), yet he still is the one who proposes a physical form (consistent with the ontology of the graph as a physical object) surrounding the medium rather than a change in the medium itself (which would represent a more obvious change in resistance for the wave).

\begin{quotation}
$\!\!\!\!\!\!\!\!\!$S2: So they're saying that there would be reflection on a potential up-step like a\ldots\newline
S1: Yeah, just like a sound, or a water wave or something.\newline
S1: Um, well 'cause I know on a potential up-step,\ldots like if you just had\ldots water and you had, you know, deeper part and a shallower part, and you had a wave, some of it would reflect back.
\end{quotation}

Here S1 applied the metaphor of a physical object again, and proposes a second analog based on the physical form of the graph rather than a change in ``density'' or ``tension'' of the medium.  Actually, a physical step on a river bed could be a valid example if S1 connected it to a model of how the resistance experienced by a surface wave attenuates with the depth of the water.  He does not, and this explains his uncertainty below.

\begin{quotation}
$\!\!\!\!\!\!\!\!\!$S1: So that's not too hard to see.  But like, I would guess that the same thing would happen if you had a down-step, but that's not something like I really, I could vouch for.   Like I think they're looking for stuff that like most people know.\newline
S2: Is that what its saying?  Its coming at it with every energy, like continuous energies, like around the step?
\end{quotation}

S2's statement is interesting.  The use of ``at'' and ``around'' are examples of grammatical {\it location} and suggest the metaphor: {\sc the step is a physical object}.  S1 shows he is still on the right track when he says:

\begin{quotation}
$\!\!\!\!\!\!\!\!\!$S1:  I think they're just asking for like, examples from\ldots in real life from when a wave\ldots goes into a space of less resistance and has reflection back.\newline
S4: So in classical what would happen at a potential down-step?\newline
S1: A potential down-step?\newline
S2: It would just keep going\ldots\newline
S1: \ldots It would just speed up. At a potential up-step it would just slow down.
\end{quotation}

\subsubsection{Discussion}

One alternative hypothesis to explain the difficulties presented above could be that the students are unable to interpret the physical meaning of the potential energy graph or are simply not understanding the situation.  However, S1's ability to interpret potential energy graphs correctly and articulate the key to the analogy discounts this hypothesis.  The data show that his inability to come up with a productive analog must be based on other factors.  Our framework explains how S1 is distracted by applying an overly literal interpretation of the {\sc potential well} metaphor in an inappropriate situation.  Possibly, a way of talking (i.e., describing the potential graph as a ``step'') is affecting students' reasoning.  Our analysis (Section~\ref{student_difficulties} above) shows that the students in this group are searching in the category of ``physical objects'' for an analogy, in accordance with the underlying ontological metaphor {\sc the potential energy graph is a physical object} rather than searching in a more productive category. Other researchers have also noticed that QM students tend to pick 2-d gravitational analogs when asked to come up with physical examples of 1-d potential energy graphs~\cite{bao1999,sadaghiani2003}.

As a control we posed the same problem to the professors in the interview study.  They all responded that an analogy of an electron beam scattering off of a potential down step is light traveling from a medium with greater index of refraction to a medium with a lesser index of refraction.  When asked why changing optical media was a good analog, most were unable to explain, but continued to elaborate their answer.  Only one professor was able to explain why this was a good analogy. Prof. E: ``I know because we've thought about these things before and its just been classified in that category.''   This statement suggests that physicists are able to automatically search for an analog in a category of analogous {\it processes} rather than analogous {\it objects}.  It may also suggest that physicists' ideas have become so tightly bound into larger conceptual units that professors are unable to break down their reasoning into smaller parts again.

We have shown how physics professors can use metaphorical systems to reason productively in certain situations while students take the same representation and apply it too literally and inappropriately in other situations.  Strange ideas like the megaphone make sense if we understand the underlying ontology of the graph, spoken of as a physical object.  We think that the example of student discourse presented above is a typical example of students' difficulties arising from linguistic representations.

\subsubsection{``Robust Misconceptions'' Related to the {\sc Potential Well} Metaphor}

Are there ``robust misconceptions'' in QM?  The characteristics of a robust misconception are that it must be (a) present before instruction, (b) common to a significant percentage of students in a particular class, and reproducible in form and structure across different classes at different institutions in different contexts, and (c) resistant to instruction.  (See~\cite{smith1993} for example.)
 
Although research on students' understanding of QM is in its infancy, it appears that students do have specific difficulties that have the characteristics of a robust misconception. One emerging example is presented in Table~\ref{tunneling_data}. It has been observed that students think that a QM particle loses energy when it tunnels through a barrier. McKagan et al., who studied this example, freely use the word ``misconception'' in their paper~\cite{mckagan2006}.

\begin{table*}[htbp]
\caption{Selected examples of the ``exhaustion'' misconception:  Summary from three studies.\label{tunneling_data}}
\begin{ruledtabular}
\begin{tabular}{p{8.5cm}p{8.5cm}}

{\bf Authors' summary and explanation} & {\bf Sample student responses used to justify this explanation.} \\
\hline
 Lei Bao~\cite{bao1999} interviewed ten students over two semesters.  Three responded with the incorrect idea that a quantum mechanical particle loses energy when it tunnels through a potential barrier.   & Bao observed that all three students gave similar responses. Mike: ``\ldots less energy so the amplitude will be reduced,\ldots Amplitude is reduced because {\it energy is lost in the passage} [our emphasis]\ldots'' \\
 \hline
 Jeffrey Morgan et al.~\cite{morgan2004} found that all six students that they interviewed thought that the particle lost energy when it went through a potential barrier. Two of the students had completed a senior level QM course and four had completed a sophomore level introductory QM course & Selena: ``Uh, because it requires energy to go through this barrier.''

Jack: ``\ldots when the particle of some \ldots energy, encounters a potential barrier, there is a possibility\ldots that a particle will actually just go straight on through, losing energy as it does so, and come out on the other side\ldots at a lower energy\ldots'' \\
\hline
  McKagan et al.~\cite{mckagan2006} gave a conceptual test to a group of engineering majors (N = 68) and  physics majors (N = 64) after they had completed a modern physics course.  One of the questions probed students' understanding of tunneling processes.  On this question 24\% of the engineering majors were able to answer correctly and 38\% of the physics majors were able to answer correctly. & No interview samples were provided, but the authors summarize the student responses as follows: ``all students who selected answers A, B, or E [more than 50\% for both engineers and physicists] argued that since energy was lost in tunneling, making the barrier wider and/or higher would lead to greater energy loss.''  \\

 \end{tabular}
 \end{ruledtabular}
 \end{table*}
 
A consistent pattern of reasoning is presented in Table~\ref{tunneling_data}. This pattern contains the following two elements: (1) It takes energy for a particle to tunnel through a barrier.  (2) Making the barrier wider or higher means that the particle loses more energy/expends more effort when tunneling through it.  Morgan et al. speculate that the difficulty may come from either (a) intuitive classical ideas about a particle passing though a barrier, or (b) physicists tend to draw the potential energy graph and the wave function superimposed.  Thus a decaying wave-function amplitude may be confused with a decrease in energy.

McKagan et al., however, noticed something interesting in their study.  In interviews, they discovered that students do not see the potential energy graph as representing the potential energy of the particle in question.  They see it rather as some external object with which the particle interacts.  The authors describe an example from their interviews:
 \begin{quotation}
$\!\!\!\!\!\!\!\!\!$``When pressed, he said that the `bump' was `the external energy that the electron interacts with' and insisted that it was not the potential energy of the electron itself, in spite of the fact that it was explicitly labeled as such in the previous question.''
 \end{quotation}
The authors speculate that statements like ``a particle in a potential'' may be the cause of this problem.

Our analysis supports this idea and provides an explanation for the underlying causes of this student difficulty.  The problem is much more widespread than just phrases like ``a particle in a potential.''  As we pointed out in Section~\ref{potential_well_analysis}, many statements that fall under the category of the {\sc potential well} metaphor, tend to separate the particle or wave function from its potential energy graph in the grammar of the sentence.  Most often the particle/wave function functions grammatically as the {\it medium} while the potential energy ``barrier'' functions as either the {\it range}, or {\it circumstance} of {\it location}.  The two grammatical {\it participants} then interact with each other by a grammatical {\it process} such as ``tunnels through'' or ``is reflected.''   We hypothesize that the language is the primary source of the students' model.  Graphical representations (such as the superposition of the energy graph and the wave-function) and classical intuitions build on and extend this basic model, leading to the idea that energy is lost in the tunneling process.

\subsubsection{Summary}

The example of the {\sc potential well} metaphor illustrates how the language used to describe certain QM systems may pose extraordinary difficulties, especially if students are not aware of how and why metaphorical terms are being used.  The metaphorical language, grounded in the classical world, may encourage students to associate extra (classical) properties to the QM system as they try to coordinate these new representations with their prior understanding of the world.  These over-extensions of the representation seem to be the source of their difficulties.

\subsection{The {\sc Bohmian} Metaphor}

As part of our study, two senior undergraduate physics majors (in their second QM course)  agreed to be videotaped while working on their QM homework together. Both were native English speakers.  In this particular session S1 and S2 were working on a problem worked out in class by the lecturer that they did not understand.  The question may be expressed as follows:  ``Given an electron in the ground state of an infinite square well of width L.  The walls are suddenly moved apart so that the width of the well becomes 2L.  What is the probability that the electron is in the ground state of the new system?''

The two students working on the problem understood the sudden approximation, they calculated the overlap integral and got a numerical answer which was reasonable.  Then S1 stopped and pondered that his answer made no sense.  He argued that his answer should be zero.  A discussion with the observer (D.T.B.) followed.

{\begin{figure}[htbp]\centering
\includegraphics[height = 3.5cm]{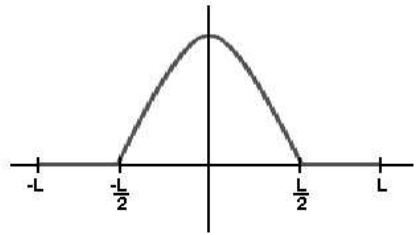}
\caption{Wave function of the electron in the sudden approximation\label{wavefn}}
\end{figure}\par}

\begin{quotation} 
$\!\!\!\!\!\!\!\!\!$S1: But I am still confused about what I was\ldots saying about if there is a probability that it is in the [sic] first ground state --- it seems to say that the particle can be where it is not.

$\!\!\!\!\!\!\!\!\!$D.T.B.: Why do you say that?

$\!\!\!\!\!\!\!\!\!$S1: Because we know that the wave function looks like this [points to a sketch similar to Fig.~\ref{wavefn}] --- Oh, so its not the probability of it being in the ground state really\ldots I think the probability is really\ldots I mean, we know that its in this state [points to sketch similar to Fig.~\ref{wavefn}] so it can't be in the ground state.  So it's zero [the probability].
\end{quotation}

The discussion circled around this theme for some time.  S1 was concerned that if the particle was ``in the ground state'' of the new well, it would permit the particle to exist outside of the \mbox{[-L/2,L/2]} region of its initial wave function.  The wave function limits where the particle can be, but to say the electron is ``in the ground state of the new well'' does not suddenly permit it to exist outside of the \mbox{[-L/2,L/2]} region; it is simply a statement about measuring the energy of the electron.   We believe that the linguistic framework we have developed provides both a reasonable and parsimonious explanation for S1's difficulties.  The prepositional phrase,``in the ground state,'' is functioning grammatically as a {\it location}.  S1's argument, that the probability should be zero, draws specifically on the location metaphor.   He says, ``it [the original question]  seems to say that the particle can be where it is not.''  This statement suggests that he is viewing the question as a question about the location of the particle.  In other words, he is interpreting the phrase ``in the ground state'' literally rather than figuratively.

This difficulty with the {\sc Bohmian} metaphor remains undocumented in the physics education research literature.  However, a physics professor who teaches undergraduate quantum mechanics, reported in a private conversation that he observed the {\it identical} difficulty amongst his students with the same sudden approximation problem.

\section{Conclusion}

We have shown that coherent systems of metaphors exist in physicists' language.  We have shown that physicists use these metaphorical systems in their language to speak and reason productively about QM systems.  They are able to invoke many different metaphorical systems, sometimes with apparently conflicting ontologies, depending on the situation they are trying to describe.   At the same time, physicists appear to understand the applicability and limitations of their metaphorical language in each situation.  We have also shown how these metaphorical systems can be identified with systematic use of both grammatical and metaphorical analysis.  And we have shown how the elaborated metaphors build on the underlying ontology encoded in the grammar.

In some cases, it seems that physicists have appropriated conceptual metaphors from language to express their ideas.  The example with Born and the {\sc Bohmian} metaphor shows how a new idea in physics comes out of a blending of older ideas into a metaphorical blend.  Likewise the final product of the language is (in this case) a blend between an analogy to Einstein's ghost field and also already existing structures in language that are normally used to describe ontological {\it mental states}.

We have presented two case studies of groups of students struggling with and being confused by overly literal interpretations of the metaphorical language they encounter in QM.  The context of QM is particularly convincing because it is difficult to argue that students enter their QM course with preconceptions or misconceptions about QM based on personal experience.  Many of the difficulties observed, appear after instruction.  It seems more plausible to hypothesize that these difficulties are related to the way in which physical ideas are presented during instruction itself.

We have presented one example (the exhaustion misconception) of a documented common conceptual  difficulty that students have with QM and how we can account for their na\"ive model with the linguistic framework we have developed.  Physicists understand that higher barrier means a slower rate of tunneling or leaking.  In contrast, students think that the particles get tired.  The underlying issue is use and misuse of the metaphorical picture.

\section{Future Directions\label{futuredirections}}

We feel that further research on the role of language in learning physics needs to examine more carefully the instructional implications of language as a legitimate representation of knowledge and ideas in physics.  For example, how can we make students more aware of the presence of physical models encoded in the metaphorical language that we use?   Can students be encouraged to think about the applicability and limitations of different metaphorical pictures~\cite{sutton1993}?  Instead of allowing students to say, ``the electron is trapped in a square well,'' unchallenged, maybe the most important question to ask students is, ``what do you mean, what is this `square well' you are talking about?''  As a corollary, maybe we should encourage students to ask us, ``what do you mean?'' when we use a metaphor such as {\sc the electron is a wave} without justifying why it is applicable and when it is not.

Does it matter how we ask questions of our students?  If we phrase a question with different grammar or different metaphors, do students respond differently?  There maybe occasions when the way in which the question is asked is obscuring the real physical understanding that students have.  There is some preliminary evidence that this may indeed be the case~\cite{schuster1983}.  

It seems to us that if we think of language as a representation and recognize its unique difficulties, we should put more effort into helping students become comfortable with this representation.  Future research could focus on student difficulties in other areas of physics that may be related to the language that students hear in the physics classroom. (See~\cite{touger1991} for example.)   In some cases difficulties may be related to linguistic models that students have developed prior to instruction.  We suggest that some student difficulties may be attempts to negotiate the meaning and applicability of different linguistic models.  Awareness of such linguistic difficulties would help teachers to facilitate their students' learning.  More research on this idea is needed.

There is one major aspect of cognitive linguistics that we have not attempted to apply to the field physics education research in this paper.   This is the idea of conceptual blending~\cite{fauconnier2002}.  Conceptual blending may provide a complementary account of many of the ideas in this paper.  Conceptual blending also has an added advantage in that it  could account for online meaning construction in terms of the blending of metaphors. This may better account for ``local'' or ``personal'' ways of expression observed among individual professors and students.  The dynamics of blending may also be useful for answering questions about how we can make students more aware of the myriad of models encoded by the metaphors in physicists' language.  We think that this may be a fruitful line of inquiry in future work.

\begin{acknowledgments}

We would like to thank the following people for their help with this paper: L. Atkins, H. Brookes, G. Horton, Y. Lin, J. Mestre, E. Redish, M. Sindel, A. Van Heuvelen, A. Zech.

\end{acknowledgments}

\bibliography{paperbibliography}

\end{document}